\documentclass[aps,amssymb,showkeys]{revtex4}
\usepackage{graphicx}
\usepackage[dvips, bookmarks, colorlinks=true, plainpages = false, citecolor = red, urlcolor = blue, filecolor = blue]{hyperref}
\usepackage{amsmath}

\newcommand{\be}{\begin{equation}}
\newcommand{\ee}{\end{equation}}
\newcommand{\bea}{\begin{eqnarray}}
\newcommand{\eea}{\end{eqnarray}}
 
\newcommand{\ba}[1]{\begin{array}{#1}}
\newcommand{\ea}{\end{array}}

\def\p{\partial} 

\newcommand{\cL}{{\cal L}}

\newcommand{\cP}{{\cal P}}

\renewcommand{\a}{\alpha}
\renewcommand{\b}{\beta}
\def\G{\Gamma}

\def\D{\Delta}
 
\def\k{\kappa}

\def\f{\phi}
\def\y{\psi}

\begin{document}

\title{General solution of an exact correlation function factorization in conformal field theory}

\author{Jacob J. H. Simmons}
\email{j.simmons1@physics.ox.ac.uk}
\affiliation{Rudolf Peierls Centre for Theoretical Physics, 1 Keble Road, Oxford OX1 3NP, UK}

\author{Peter Kleban}
\email{kleban@maine.edu}
\affiliation{LASST and Department of Physics \& Astronomy,
University of Maine, Orono, ME 04469, USA}

\date{\today}
\begin{abstract}
The correlation function factorization 
\be \nonumber
\langle \phi(x_1) \phi(x_2) \psi(z, \bar z)\rangle = K \sqrt{\langle \phi(x_1) \psi(z, \bar z)\rangle \langle \phi(x_2) \psi(z, \bar z)\rangle \langle \phi(x_1) \phi(x_2) \rangle} \; ,
\ee
with $K$ a boundary operator product expansion coefficient, is known to hold for certain scaling operators %
 at the two-dimensional percolation point %
 and in a few other cases.  Here the correlation functions are evaluated in the upper half-plane (or any conformally equivalent region) with $x_1$ and $x_2$ arbitrary points on the real axis, and $z$ an arbitrary point 
 in the interior.  

 This type of result is of interest because it is both exact and universal, relates higher-order correlation functions to lower-order ones, and has a simple interpretation in terms of cluster 
 or loop %
 probabilities in several statistical models.  
   
 This motivated us to use the techniques of conformal field theory to determine the general conditions for its validity.

 Here, we discover that either  $\langle \phi_{1,3}(x_1) \phi_{1,3}(x_2) \phi_{1/2,0}(z, \bar z)\rangle$ or %
  $\langle \phi_{3,1}(x_1) \phi_{3,1}(x_2) \phi_{0,1/2}(z, \bar z)\rangle$
 factorizes in this way for any central charge $c$, generalizing previous results. In particular, the factorization holds for either FK (Fortuin-Kasteleyn) or spin clusters in the $Q$-state Potts models; it also applies to either the dense or dilute phases of the $O(n)$ loop models.
 
 Further, only one other non-trivial set of highest-weight operators (in an irreducible Verma module) factorizes in this way.  In this case the operators have negative dimension (for $c < 1$) and do not seem to have a physical realization.  %
\end{abstract}
\maketitle

\section{Introduction} \label{intro}

Usefully characterizing  fluids at thermal equilibrium requires knowledge of correlation functions.  Many expressions for important experimental and theoretical quantities involve these functions.  Higher-order correlations of quantities such as the density $\rho(x)$ at several points, e.g.\   $\langle \rho(x_1)\rho(x_2)\rho(x_3)...\rangle$ (where the brackets denote a thermal average) are ubiquitous but especially difficult to calculate.  A variety of approaches have been proposed (for a review, see \cite{St}).  One idea is to factorize the higher-order correlations in terms of lower-order correlations (with fewer points), since the latter are generally better known.   

Recent work, using conformal field theory,  exhibited several exact formulas in which three-point correlations factorize in terms of two-point correlations or correlations involving one point and an interval \cite{KSZ,SZK}.  Related results (involving one point and two intervals) give factorizations that are not quite exact, but with small corrections \cite{KSZ2}. To our knowledge, these are the only such exact (or almost exact) results for interacting fluids of any type.  These formulas apply to percolation in two dimensions at the percolation point in the upper half-plane (or any simply connected region) \cite{KSZ,KSZ2,SZK}, the critical $Q$-state Potts models \cite{KSSZ}, and critical percolation in three dimensions \cite{ZKS}.  

\begin{figure}[ht]
\begin{center}
\includegraphics[width=2.0in]{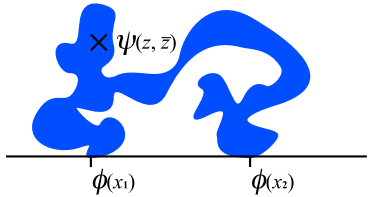}
 \end{center}
\caption{(color online) Cluster connections for factorization.} \label{fig1}
\end{figure}

One of these exact factorizations, first found in \cite{KSZ} for a system exhibiting critical percolation, may be understood in terms of the probability that various points are connected %
via  %
 clusters, as shown in Figure \ref{fig1}  %
 (for a cluster connecting all three points).  %
 Here 
\be \label{probfact}
\cP(x_1,x_2,z)= K \sqrt{\cP(x_1,x_2) \cP(x_1,z) \cP(x_2,z)} \; ,
\ee
where $x_1$ and $x_2$ are distinct boundary points (e.g.\ on the real axis), $z$ lies on the boundary or in the interior, and $\cP(\cdot)$ is the probability that the given points lie in the same 
cluster. The result holds in any region conformally equivalent to the upper half-plane, with $K$ a universal constant.  

The factorization in (\ref{probfact}) superficially resembles the Kirkwood superposition approximation \cite{K}.  However, the formulas are not the same, and (\ref{probfact}) is both exact and universal, while there does not seem to be any interacting fluid for which the Kirkwood formula  applies exactly.

The  result (\ref{probfact}), for critical percolation, %
 corresponds to a $c=0$ CFT correlation function with boundary and bulk magnetization operators $\f=\f_{1,3}$ \cite{JC84} and $\psi=\f_{1/2,0}$  \cite{MdN83}.   In this case the constant $K$ has been evaluated \cite{SZK,SK}, giving
 \bea \label{const}
 K &=& \frac{2^{7/2}\;\pi^{5/2}}{3^{3/4}\;\G(1/3)^{9/2}}  \nonumber\\
 &=& 1.0299268 \ldots \; ,
 \eea
and   
 both (\ref{probfact}) and (\ref{const}) agree with high-precision numerical simulations \cite{KSZ}.    Other factorizations are also known to agree well with numerical results \cite{KSZ,KSZ2,KSSZ,ZKS}.

In terms of conformal quantities, (\ref{probfact}) can be expressed as
\be \label{fact}
\langle \phi(x_1) \phi(x_2) \psi(z, \bar z)\rangle = K \sqrt{\langle \phi(x_1) \psi(z, \bar z)\rangle \, \langle \phi(x_2) \psi(z, \bar z)\rangle \, \langle \phi(x_1) \phi(x_2) \rangle} \; ,
\ee
where $\phi$ and $\psi$ are appropriately chosen boundary and bulk 
primary 
operators, respectively, and $K$ is a (boundary) operator product expansion coefficient.

In this paper, we use conformal field theory to find all choices of $\f$ and $\y$ for which the factorization (\ref{fact}) applies.  %
   We show that there are only three non-trivial combinations of highest-weight operators (of irreducible Verma modules) that factorize in this way.  In each case  %
   (\ref{fact})  %
 holds for general central charge. 
Operators that can be identified for continuous central charge are given Kac indices that do not change with $c$, as is common for continuous parameter loop models. Thus the bulk magnetization operator mentioned above is denoted $\f_{1/2,0}$, even though neither index is a positive integer.

The percolation model is equivalent to the FK random cluster representation of the $Q=1$ Potts model. %
The  equivalent correlation functions for the $0 \le Q \le 4$ Potts models  factorize in the same way  %
 \cite{KSSZ}. Explicit values of $K$ are known for these cases \cite{SK}, though the probabilistic interpretation similar to (\ref{probfact}) is slightly more complicated.  Thus  %
this factorization extends to  FK clusters in Potts models  %
  with $0 \le Q \le 4$.

The factorization also applies when $\phi = \phi_{3,1}$ and $\psi = \phi_{0,1/2}$.  These weights are Potts model exponents at the tricritical point.  
There is a tricritical $Q$-state Potts model that encodes information about the geometrical clusters of the  
Potts model, but at a different $Q$ value \cite{JS04} (tricritical $Q=1$ corresponds to geometric $Q=2$, for example).  %
Thus this second solution implies that the geometric clusters (blocks of like-spin in the $Q \in \mathbb{Z}$ case) obey the same factorization as the FK clusters.  From the  Schramm-Loewner Evolution (SLE) point of view, the  FK cluster results are dual to the geometric cluster results under $\k \to 16/\k$, where $\k$ is the SLE parameter.  %
Using the above, we see that  %
  the previous result \cite{KSZ} (at $\k = 6$) extends to $8/3 \le \k \le 8$.   Note that the %
connection  %
 between FK and geometric clusters is identical to that between the dense and dilute phases of the $O(n)$ model.
   
   Finally, the factorization holds for $\phi = \phi_{1/3,1/3}$ and $\psi = \phi_{2/3,2/3}$. Here the dimensions of  both operators are  negative for $c < 1$.  These operators have a simple explanation in the vertex operator formalism, but  we are not aware of any physical realization.  

\section{Conditions on the Conformal Weights}

In this section we  make use of the operator product expansion to determine necessary conditions on the conformal weights for factorization. We then solve the conditions, which 
is sufficient to determine all cases for which (\ref{fact}) is valid.

 The operators in (\ref{fact}) satisfy
\bea
\label{bdryOPE}
\f(x)\f(x+\Delta x) &=& \sum_{q \in S} C_{\f \f q}\; \Delta x^{h_q-2 h_{\f}} \sum_{\{ \ell \}} \Delta x^{| \ell |} \b_{\{ \ell \}} L_{-\{ \ell \}}q(x)\; , \quad \mathrm{and}\\
\label{bulkOPE}
\y(z,\bar z) &=& \sum_{p \in R} C_{\y; p}\; y^{h_p-2 h_{\y}} \sum_{\{ \ell \}} y^{| \ell |} \b_{\{ \ell \}} L_{-\{ \ell \}}p(x)\; ,
\eea
where $S$ is the set of all operators appearing in the boundary-boundary fusion rules, and $R$ is the set of all operators appearing in the bulk-boundary fusion rules.  We assume that the boundary fields $q$ and $p$ on the right hand side of these relations are elements of standard 
irreducible Verma modules so that the normal scheme of generating descendants via the Virasoro generators applies. 
Note that this assumption 
excludes logarithmic modules. However, if the operators were logarithmic, the right hand side of (\ref{fact}) would generate $\log^{1/2}(\cdot)$ factors, which are incompatible with the general form of a logarithmic four-point function.

Now the orthogonality of conformal families 
 ensures 
that the only bulk-boundary fusion channel that leads to non-zero three-point functions $\langle \phi(x_i)  \psi(z, \bar z)\rangle$ on the right-hand side of (\ref{fact}) is the $\f$ channel.  Similarly,
only channels common to both (\ref{bdryOPE}) and (\ref{bulkOPE}) will contribute to the left hand side of 
(\ref{fact}); i.e. 
 only channels in $S \cap R$ will contribute to the four-point function $\langle \phi(x_1) \phi(x_2) \psi(z, \bar z)\rangle$.

 Thus a necessary condition for the factorization is $\f \in S \cap R$.  
Hence only the $\f$ channel remains on the right hand side, and if one expands it about $y=0$, this determines the powers of $y$ that enter.  It follows that  only that channel can contribute to the left hand side (otherwise different powers of $y$ would appear). Thus,  we have $\f = S \cap R$, i.e. 
 only the $\phi$ channel can be common to the two operator product expansions.  This condition can also be expressed in terms of operator product expansion coefficients as
\be 
\f = Q \cap P \leftrightarrow C_{ \f \f p}C_{\y; p} \propto  \delta(p,\f)\; . 
\ee
This is a  strong condition on the permissible fusion rules, as we will see.  

Limiting the calculation to  the $\phi$ channel only 
 gives 
\be
\label{bdryOPE2}
\f(x)\f(x+\Delta x) = C_{\f \f \f}\; \Delta x^{-h_{\f}} \sum_{\{ \ell \}} \Delta x^{| \ell |} \b_{\{ \ell \}} L_{-\{ \ell \}}\f(x)\; , \quad 
\ee 
which we insert into the left hand side of the factorization (\ref{fact}) 
giving
\be\label{exp}
\langle \f(x_1)\f(x_2) \y(z,\bar z) \rangle= C_{\f \f \f}\; \D x^{-h_{\f}} \sum_{\{ \ell \}} \D x^{| \ell |} \b_{\{ \ell \}} \cL_{-\{ \ell \}}^{x_1}\langle \f(x_1) \y(z, \bar z)\rangle\; ,
\ee
where 
\be \label{Lop}
\cL_{-\ell }^x = \sum_{i}\left( \frac{(\ell-1) h_i}{(z_i- x)^{\ell}}-\frac{\p_{z_i}}{(z_i- x)^{\ell-1}} \right)=\frac{(\ell-1) h_i}{(z- x)^{\ell}}-\frac{\p_{z}}{(z- x)^{\ell-1}}+\frac{(\ell-1) h_i}{(\bar z- x)^{\ell}}-\frac{\p_{\bar z}}{(\bar z- x)^{\ell-1}}
\ee 
 is the standard operator arising from commuting $L_{-\{ \ell \} }$ over all the fields in the correlator and $\D x := x_2-x_1$.

The form of the two point function in expression (\ref{exp}) is fixed by conformal symmetry
\be\label{bb2pt}
\langle \f(x_1) \y(z, \bar z)\rangle = C_{\y; \f} \a_{\f} \left(\frac{z-\bar z}{2 i}\right)^{h_{\f}-2 h_{\y}}\left( \frac{1}{(z - x_1)(\bar z - x_1)} 
\right)^{h_{\f}}\; ,
\ee
where $\a_{\f}$ encodes  the boundary operator normalization \cite{L}. Inserting the explicit form of 
(\ref{Lop}) 
 in (\ref{exp}) and making use of (\ref{bb2pt})  
 gives, for the left hand side of the (\ref{fact}), 
\bea \nonumber
\langle \f(x_1)\f(x_2) \y(z,\bar z) \rangle
\nonumber&=& C_{\f \f \f}\; \D x^{-h_{\f}}\langle \f(x_1) \y(z, \bar z)\rangle \left(1
+  \D x\;  \b_1 \frac{\p_{x_1}\langle \f(x_1) \y(z, \bar z)\rangle}{\langle \f(x_1) \y(z, \bar z)\rangle} \ldots\right. \\
&\nonumber&\left. \phantom{C_{\f \f \f}\; \D x^{-h_{\f}}} +  \D x^2 \b_{1 1} \frac{\p_{x_1}^2\langle \f(x_1) \y(z, \bar z)\rangle}{\langle \f(x_1) \y(z, \bar z)\rangle}
+  \D x^2 \b_2 \frac{\cL_{-2}^{x_1}\langle \f(x_1) \y(z, \bar z)\rangle}{\langle \f(x_1) \y(z, \bar z)\rangle} + \mathcal{O}(\D x)^3\right)\\
\label{lhsexp}&=& C_{\f \f \f}\; \D x^{-h_{\f}}\langle \f(x_1) \y(z, \bar z)\rangle \left(1
+  \D x\;  \b_1 h_{\f}  \left( \frac{1}{z-x_1}+\frac{1}{\bar z - x_1}\right) \ldots\right. \\
&\nonumber&\phantom{C_{\f \f \f}\; \D x^{-h_{\f}}}
+  \D x^2 \b_{1 1} \left( \frac{h_{\f}(h_{\f}+1)}{(z-x_1)^2}+\frac{h_{\f}(h_{\f}+1)}{(\bar z - x_1)^2}+\frac{2h_{\f}{}^2}{(z-x_1)(\bar z-x_1)} \right)\ldots\\
&\nonumber& \left. \phantom{C_{\f \f \f}\; \D x^{-h_{\f}}}
+ \D x^2 \b_2 \left( \frac{h_{\f}+h_{\y}}{(z-x_1)^2}+\frac{h_{\f}+h_{\y}}{(\bar z - x_1)^2}+\frac{h_{\f}-2 h_{\y}}{(z-x_1)(\bar z-x_1)} \right) + \mathcal{O}(\D x)^3\right)\; ,
\eea
where  $\cL_{-1}^x = \p_x$ was used to simplify the expressions.

Now consider the right hand side of (\ref{fact}).  One of the correlation functions appearing there is given   
in (\ref{bb2pt}) 
and the others are
\bea \label{bB2pt}
\langle \f(x_1) \f(x_2) \rangle&=&\a_{\f} \D x^{-2 h_{\f}} \; ,\\ \nonumber
\langle \f(x_2) \y(z, \bar z)\rangle&=& C_{\y; \f} \a_{\f} \left(\frac{z-\bar z}{2 i}\right)^{h_{\f}-2 h_{\y}}\left( \frac{1}{(z - x_2)(\bar z - x_2)}\right)^{h_{\f}}\\
\nonumber&=& C_{\y; \f} \a_{\f} \left(\frac{z-\bar z}{2 i}\right)^{h_{\f}-2 h_{\y}}\left( \frac{1}{(z - x_1-\D x)(\bar z - x_1-\D x)}\right)^{h_{\f}}\\
&=&\langle \f(x_1) \y(z, \bar z)\rangle \left(1-\frac{\D x}{z - x_1}\right)^{-h_{\f}}\left(1-\frac{\D x}{\bar z - x_1}\right)^{-h_{\f}}\; . \label{bb2pt2}
\eea
Combining all three gives an exact expression for the right hand side of (\ref{fact}):
\be \label{rhsexp}
K \sqrt{\langle \phi(x_1) \psi(z, \bar z)\rangle \langle \phi(x_2) \psi(z, \bar z)\rangle \langle \phi(x_1) \phi(x_2) \rangle}=\frac{K \sqrt{\a_{\f}} \D x^{-h_{\f}}\langle \f(x_1) \y(z, \bar z)\rangle}{\left(1-\frac{\D x}{z - x_1}\right)^{h_{\f}/2}\left(1-\frac{\D x}{\bar z - x_1}\right)^{h_{\f}/2}}\; .
\ee
Comparing the  terms of ${\cal O}(\Delta x^0)$ in (\ref{lhsexp}) and (\ref{rhsexp}) we find that $K=C_{\f \f \f} \a_{\f}{}^{-1/2}$.  Comparing at ${\cal O}(\Delta x^1)$ and ${\cal O}(\Delta x^2)$ lead, respectively, to

\bea
\frac{1}{2} h_{\f}  \left( \frac{1}{z-x_1}+\frac{1}{\bar z - x_1}\right)&=&\b_1 h_{\f}  \left( \frac{1}{z-x_1}+\frac{1}{\bar z - x_1}\right)\\
\frac{1}{8}\left( \frac{h_{\f}(2+h_{\f})}{(z-x_1)^2}+\frac{h_{\f}(2+h_{\f})}{(\bar z - x_1)^2}+\frac{2 h_{\f}{}^2}{(z-x_1)(\bar z-x_1)}\right)
&=&\b_{1 1} \left( \frac{h_{\f}(h_{\f}+1)}{(z-x_1)^2}+\frac{h_{\f}(h_{\f}+1)}{(\bar z - x_1)^2}+\frac{2h_{\f}{}^2}{(z-x_1)(\bar z-x_1)} \right)\\
&\nonumber& \quad + \b_2 \left( \frac{h_{\f}+h_{\y}}{(z-x_1)^2}+\frac{h_{\f}+h_{\y}}{(\bar z - x_1)^2}+\frac{h_{\f}-2 h_{\y}}{(z-x_1)(\bar z-x_1)}\right)\; ,
\eea
 
giving
\bea
1/2&=&\b_1\; ,\\ \label{JScond1}
h_{\f}(2+h_{\f})&=&8\b_{1 1} h_{\f}(1+h_{\f})+8\b_2 (h_{\f}+h_{\y})\; \mathrm{,\; and}\\\label{JScond2}
h_{\f}{}^2&=&8\b_{1 1}h_{\f}^2+4\b_2(h_{\f}-2 h_{\y})\; ,
\eea
as necessary conditions for the factorization.  

On the other hand, 
we can directly determine the beta coefficients   
in 
(\ref{bdryOPE2}) 
using the methods of 
\cite{BPZ},  giving 
\bea
1/2 &=& \b_1\; ,\\  \label{BPZcond1}
(1+h_{\f}) &=& 4(1+2 h_{\f})\b_{1 1} + 6 \b_2\quad \mathrm{and},\\ \label{BPZcond2}
4 h_{\f} &=& 12 h_{\f} \b_{1 1}+(8h_{\f}+c)\b_2\; ,
\eea
where $c$ is the central charge.  
This analysis is valid as long as $\f$ is a regular operator whose descendants have a standard 
irreducible Verma module structure. 
This  assumption was 
 already made above, so there is no additional loss of generality. 

 (\ref{BPZcond1}) and (\ref{BPZcond2}) may be inverted as long as 
  $h_\f \neq h_{1,2}, \;h_{2,1}$.  
  (The condition reflects the fact that 
   these two operators have second order null vectors.) However for these operators $[\f] \notin [\f] \times [\f]$ so they are not of interest to us. 
 For any other case we find  

\bea
\b_{1 1} &=&\frac{ c (1 + h_{\f})+8 h_{\f}( h_{\f}-2)}{4 c (1+ 2 h_{\f})+8 h_{\f}(8h_{\f}-5)}\; ,\\
\b_2 &=&\frac{h_{\f}(1 + 5 h_{\f})}{ c (1+ 2 h_{\f})+2 h_{\f}(8h_{\f}-5)} \; .
\eea
Inserting these expressions for $\b_{1 1}$ and $\b_2$ into (\ref{JScond1}) gives either
\be \label{mineq}
h_{\f} = 0  \; ,
\ee
or
\be \label{majeq}
 h_{\y}=\frac{h_{\f}(4+c-2h_{\f})}{8(1+5 h_{\f})} \; .
\ee
(Note that (\ref{mineq}) or (\ref{majeq})  satisfies (\ref{JScond2}) as well.)  Either one is a necessary (but not sufficient) condition for the factorization (\ref{fact}) to hold.  Note that if, following (\ref{mineq}), we select the identity for $\f$ the factorization becomes trivial.  

Continuing the above analysis through terms of ${\cal O}(\Delta x^3)$, we find no new conditions.  At ${\cal O}(\Delta x^4)$, three equations emerge.  Since they are rather long we refrain from displaying them.  However, making use of (\ref{majeq}) in each of them in order to re-express $ h_{\y}$ in terms of $h_{\f}$ gives rise to the single new condition
\be \label{4thorder}
\frac{h_{\f} (3 h_{\f}+2) (c-27 h_{\f}-1) \left((c-7) h_{\f}+c+3 h_{\f}^2+2\right)}{(5 h_{\f}+1) (c+8 h_{\f}-1) (5 c (2 h_{\f}+3)+2 (h_{\f}-1) (8 h_{\f}-33))}=0 \; .
\ee

There are five solutions of (\ref{4thorder}), 
but the apparent solution $h_\f=-2/3$ fails if we continue the analysis to ${\cal O}(\Delta x^6)$.  
We next examine the remaining four solutions.  

First, $ h_{\f} = 0$, which allows any value for $h_{\y}$.  This is the trivial solution with $\f$ equal to the identity. It obeys the factorization with $K=1$.  

Next is 
 $2 + c - (7 - c) h + 3 h^2 =0$, solved by the dimensions of the Kac operators $\f_{1,3}$ and $\f_{3,1}$.  By (\ref{majeq}) the corresponding bulk operators have dimensions $h_{\y}=h_{0,1/2}$ and  $h_{1/2,0}$ respectively.  As we mentioned in the introduction these two pairs $h_\f$ and $h_\y$ are critical and tricritical Potts model exponents.

One might question these solutions,  since the analysis at ${\cal O}(\Delta x^4)$ is singular for both $\f=\f_{1,3}$ and $\f_{3,1}$, due to level three null descendants in these conformal families.  However, it is straightforward to explicitly construct the differential equation associated to the third order null state  %
and verify the factorization directly  (as  in \cite{KSZ} and \cite{KSSZ}).  %

   These two correlation functions generalize the results of \cite{KSZ}, which considered percolation only.  The appearance of both critical and tricritical exponents means that the factorization (\ref{fact}) holds for either FK clusters or geometric clusters in the  $Q$-state Potts models for $0 \le Q \le 4$.  Note that in the former case, the boundary conditions are free (except at the points $x_1$ and  $x_2$), while for the latter they are more complicated (the $3$-state Potts model, for example, is discussed in \cite{GC}).
  
The factorization also has an interpretation in the $O(n)$ loop model.  Here, the geometric magnetic correlation between two boundary and one bulk points is measured,  
\emph{i.e.}~whether there is a path connecting all three points that does not cross an $O(n)$ loop.  The two sets of  
correlation functions for each central charge represent the dense phase (corresponding to the FK clusters) or 
the dilute phase (corresponding to the geometric spin clusters).  For a given central charge these two phases are dual in the SLE sense.   The  
factorization  holding  %
for dual sets of operators (where the order of all Kac indices in the correlation function are reversed, or equivalently, 
the SLE parameter $\k \to 16/\k$)  %
therefore implies that it applies across the SLE or conformal loop ensemble %
transition from dilute to dense loops that occurs at $\k=4$.  In this sense these two sets of Kac operators constitute a single physical factorization valid for all $8/3 \le \k \le 8$.

The remaining solution contains the boundary dimension, $h_{\f} = -(1-c)/27=h_{1/3,1/3}$ which  
  implies the bulk conformal weight, $h_{\y}=- 5(1-c)/216=h_{2/3,2/3}$.  
  Both of these are negative for $c < 1$.  
  
    These weights are unique as Coulomb gas vertex operators because they are the only weights which satisfy the conditions for this factorization at arbitrary central charge without screening operators.  
By contrast,   the case discussed above with $h_{\f}=h_{1,3}$ and $h_{\y}=h_{1/2,0}$ requires two screening operators.  The absence of screening charges means that this collection of operators will satisfy the factorization in question exactly for arbitrary central charge.  However, it is not immediately apparent what these operators may represent in terms of cluster or loop degrees of freedom through Coulomb gas or SLE formalisms.

Finally, note that  
 the analysis performed to ${\cal O}(\Delta x^4)$ restricts the boundary conformal weight to one of five values, of which four (including the trivial case of the identity operator) survive to higher order.  This manipulation is valid only for conformal families with standard Verma module structure up to and including level four.  %
 Therefore  
  Kac operators with null states at levels one through four could conceivably  factorize without satisfying the condition (\ref{4thorder}).  However, there are only eight operators of this type, and of these only $\f_{1,1}$,  $\f_{1,3}$, and $\f_{3,1}$, 
  already considered, obey the fusion constraint $[\f] \times [\f] = [\f]$.  Hence  no new sets of operators arise.

\section{Conclusions and Discussion}

In this note, in the context of conformal field theory, we investigate the uniqueness  of the  factorization (\ref{fact}) originally discovered in the case of critical percolation.  We restrict our attention to highest weight operators  in a standard Verma module.  Thus logarithmic modules are not explicitly considered (however, we do not expect correlation functions involving logarithmic operators to be consistent with (\ref{fact}) in general).  Using techniques of conformal field theory, we show that only three non-trivial correlation functions of conformal operators can factorize in this way.  They are $\langle  \f_{1,3}(x_1)\f_{1,3}(x_2)\f_{1/2,0}(z,\bar z) \rangle$, $\langle  \f_{3,1}(x_1)\f_{3,1}(x_2)\f_{0,1/2}(z,\bar z) \rangle$, and $\langle  \f_{1/3,1/3}(x_1)\f_{1/3,1/3}(x_2)\f_{2/3,2/3}(z,\bar z) \rangle$.  These factorizations hold for arbitrary central charge.  The first two apply to FK or spin cluster connection probabilities in the $Q$-state Potts models, respectively, and to both phases of the $O(n)$ model.  The last one has negative weights for $c<1$ and, to our knowledge, no physical realization.    

Note that one important condition for factorization is the existence of algebraic solutions for the correlation functions.  Such solutions occur in conformal field theory in various cases.  Some of these give rise to factorizations different from (\ref{fact}), see for instance \cite{SZK}.  It is probable that some of these can be generalized as well.  However,  in this work we have restricted our attention to  finding all instances of (\ref{fact}), since to our knowledge it is the simplest  exact factorization. 

Of course, our results do not address the interesting question of {\it why} the factorization is valid, or whether there is a single explanation that holds for the various different physical models to which it applies.

It may prove interesting to
consider factorization in higher dimensions, but the machinery
of CFT is considerably weaker in dimensions greater than two and different
approaches would be necessary. We plan to explore factorization of the  bulk-boundary-boundary
 three point correlation in critical percolation in three dimensions, along with
numerical simulations, in future work \cite{ZKS}.

 \section{Acknowledgments}
 
This work was supported in part by EPSRC Grant No. EP/D070643/1 (JJHS) and by the National Science Foundation Grant No.  DMR-0536927 (PK).



\begin{thebibliography}{99}

 \bibitem{St}
 G. Stell, in {\it The equilibrium theory of classical fluids},  H. L. Frisch and J. L. Lebowitz, eds. (New York: W. A. Benjamin) 1964.

 \bibitem{KSZ}
  Peter Kleban,  Jacob J. H.  Simmons, and Robert M. Ziff,  {\it Anchored critical percolation clusters and 2D electrostatics},
 Phys. Rev. Lett.
  {\bf 97} 115702 (2006) \href{http://arxiv.org/abs/cond-mat/0605120}{[arXiv: cond-mat/0605120]}.
    
   \bibitem{SZK}
  Jacob J. H.  Simmons, Peter Kleban,   and Robert M. Ziff, {\it Exact factorization of correlation functions in two-dimensional critical percolation}, Phys. Rev. E  {\bf 76} 041106, 2007,  \href{http://arxiv.org/abs/0706.4105}{[arXiv:0706.4105]}.
    
  \bibitem{KSZ2}
  Jacob J. H.  Simmons, Robert M. Ziff, and Peter Kleban, {\it  Factorization of percolation density correlation functions for clusters touching the sides of a rectangle}, J. Stat. Mech. (2009) P02067   doi: 10.1088/1742-5468/2009/02/P02067,   \href{http://arxiv.org/abs/0811.3080}{[arXiv:0811.3080]}.   

   \bibitem{KSSZ}
  Peter Kleban,  Jacob J. H.  Simmons, Thomas Stone and Robert M. Ziff,  {\it in preparation}.
  
   \bibitem{ZKS}
 Robert M. Ziff, Peter Kleban, and  Jacob J. H.  Simmons,   {\it in preparation}.
 
 \bibitem{SK}
Jacob J. H.  Simmons and  Peter Kleban, {\it First column boundary operator product expansion coefficients}, {\it  preprint},  \href{http://arxiv.org/abs/cond-mat/0605120}{[arXiv:0712.3575]}.

\bibitem{K}
J. G. Kirkwood, {\it Statistical mechanics of fluid mixtures}, J. Chem. Phys. {\bf3} 300-313 (1935).

\bibitem{JC84}
J.~L.~Cardy, {\it Conformal invariance and surface critical behavior}, Nucl. Phys. {\bf B240} 524, 1984.

\bibitem{MdN83}
M.~den Nijs, {\it Extended scaling relations for the magnetic critical exponents of the Potts model}, Phys. Rev. B {\bf 27} 1974, 1983.

\bibitem{JS04}
W.~Janke and A.~Schakel, {\it Geometrical vs. Fortuin-Kasteleyn clusters in the two-dimensional q-state Potts model},  Nucl. Phys. {\bf B700} 385, 2004 [arXiv:cond-mat/0311624].

 \bibitem {L}
 David C. Lewellen, {\it Sewing constraints for conformal field theories on surfaces with boundaries}, Nuc. Phys. B {\bf 372},  654Ñ682 (1992).

 \bibitem{BPZ}
A. A. Belavin, A. M. Polyakov, and A. B. Zamolodchikov, {\it Infinite conformal symmetry in two-dimensional quantum field theory}, Nucl. Phys. {\bf B241}, 333-380 (1984).

 \bibitem{GC}
Adam Gamsa and John Cardy, {\it SLE in the three-state Potts model - a numerical study}, J. Stat. Mech. (2007) P08020 [arXiv:0705.1510].


\end{thebibliography}
\end{document}